# A current-voltage model for Schottky-barrier graphene based transistors

David Jiménez

Departament d'Enginyeria Electrònica, Escola Tècnica Superior d'Enginyeria,
Universitat Autònoma de Barcelona, 08193-Bellaterra, Barcelona, Spain.

Corresponding author: david.jimenez@uab.es

**Abstract-** A low complexity computational model of the current-voltage characteristics for graphene nano-ribbon (GNR) field effect transistors (FET), being able to simulate a hundred of points in few seconds using a personal computer, is presented. For quantum capacitance controlled devices, self-consistent calculations of the electrostatic potential can be skipped. Instead, analytical closed-form electrostatic potential from Laplace's equation yields accurate results compared with that obtained by self-consistent Non-Equilibrium Green's Functions (NEGF) method. The model includes both tunnelling current through the Schottky barrier (SB) at the contact interfaces and thermionic current above the barrier, properly capturing the effect of arbitrary physical and electrical parameters.

## 1. Introduction

Graphene has recently emerged as a potential candidate material for nanoelectronics due to its electronic properties [1]. Geometrically is a monolayer of carbon atoms tightly packed into a 2D honeycomb lattice known to be a zero-gap material that could be fabricated using mechanical exfoliation [2] and epitaxial growth [3]. Interestingly, graphene could be patterned in nano-ribbons, using planar technologies as electron beam lithography and etching [3,4], having properties theoretically predicted to range from metallic to semiconducting depending on their width and edges [5]. This band-gap tuning capability and the possibility of large-scale integration using planar technologies open a route towards an all-graphene electronic nanodevices and circuits. Notably, recent studies [2] reported mobilities for electrons and holes in graphene of the order of $10^4$ cm$^2$/V·s. However, mobility for GNRs is expected to have smaller values than graphene, with an inverse dependence with the band gap [6], but conclusive experimental studies still lack. At this early state of development of GNR technology it seems timely to develop models of building blocks helping to conduct experiments in line with previously reported models for carbon nanotube based devices [7,8]. This work presents a simple model to analyze or design the current-voltage (I-V) characteristics of GNR-FETs as a function of physical parameters, such as GNR width (W) and gate insulator thickness ($t_{ins}$), and electrical parameters, such as SB height ($\varphi_{SB}$). This approach prevents the computational burden that self-consistency implies by using a closed-form electrostatic potential from Laplace's equation. This simplification yields accurate results compared with self-consistent results from NEGF's method [9] for the relevant limit dominated by the GNR quantum capacitance [10] ($C_{GNR}$). Note that it appears to be the interesting case for advanced applications because the ability of the gate to control the potential in the channel is maximized.

## 2. Graphene nano-ribbon electrostatics

Let us consider a semiconducting GNR contacted with metal electrodes acting as source/drain (S/D) reservoirs [Figs. 1(a)-(b)]. The resulting spatial band diagram along the transport direction has been sketched in Fig. 1(c). For a long-channel transistor the potential energy at the central region is exclusively controlled by the gate electrode and I further assume that: (i) $C_{GNR}$ dominates the total gate capacitance $C_G^{-1} = C_{ins}^{-1} + C_{GNR}^{-1} \approx C_{GNR}^{-1}$, where $C_{ins}$ represents the geometrical capacitance; and (ii) $C_{GNR} \approx 0$. The validity of the latter assumption depends on the quantum confinement strength. Downscaling W produces an increasing separation between adjacent peaks of the density-of-states versus energy, being more difficult to induce mobile charge (Q) into the GNR for reasonable values of gate voltage, meaning that $C_{GNR} = dQ/d\varphi_S \to 0$ (being $\varphi_S$ the surface potential). In the quantum capacitance limit the problem can be highly simplified because the electrostatic is governed by Laplace's equation, instead of the more involved Poisson's equation, having two important consequences affecting the band diagram: (i) the central region shifts following the gate voltage in a 1:1 ratio or, equivalently, $\varphi_S=V_{GS}$; and (ii) the band edge near the contact region has a simple analytical closed-form. For instance, the conduction band edge potential energy can be written as:

$$E_C(z) = \varphi_{SB} - \frac{2V_{GS}}{\pi}\arccos\left(e^{\frac{-z\pi}{2t_{ins}}}\right), \quad 0 < z < \frac{L}{2}$$

$$E_C(z) = (\varphi_{SB} - V_{DS}) - \frac{2(V_{GS}-V_{DS})}{\pi}\arccos\left(e^{\frac{(z-L)\pi}{2t_{ins}}}\right), \quad \frac{L}{2} \leq z < L \quad (1)$$

where L is the channel length. This expression applies to a double-gate planar geometry in the long-channel limit with vanishing contact thickness [11] [Fig. 1(a)]. The valence band can be written as $E_V(z)=E_C(z)-E_g$, where $E_g$ is the energy gap. Analytical expressions of $E_g$ for armchair shaped edges GNRs with arbitrary chirality have been derived by Son et al. [12], presenting an inverse dependence with W.

An interesting question is why self-consistency is not needed for quantum capacitance controlled devices. A simple model for self-consistency shows that the actual channel potential (U) is intermediate between the Laplace potential ($U_L$) and the potential needed to keep the channel neutral ($U_N$) [13]. Self-consistency means to determine $U_N$ and U simultaneously. In the quantum capacitance limit, the channel potential is simply $U_L$ and we can skip self-consistency. Next, I address the question about the design window where the quantum capacitance limit is relevant. The regime dominated by the quantum capacitance fulfils the condition $C_{GNR}<C_{ins}$, where

$$C_{ins} = N_G \varepsilon \varepsilon_0 \left(\frac{W}{t_{ins}} + \alpha\right) \quad (2)$$

$N_G$ refers to the number of gates, $\varepsilon$ is the insulator relative dielectric constant, and $\alpha$ is a fitting parameter $\approx 1$ [10]. Hence, the quantum capacitance dominates the gate capacitance as long as:

$$N_G \varepsilon \left( \frac{W}{t_{ins}} + \alpha \right) > \frac{C_{GNR}}{\varepsilon_0} \quad (3).$$

Using for instance the quantum capacitance ($C_{GNR}$) value for a nano-ribbon with W=5 nm, of about 10 pF/cm [10], the combination $N_G$=2, $k$=16, and $t_{ins}$=2 nm just fulfils the above inequality, meaning that the quantum capacitance limit should be relevant, in general, for low thickness and high-k insulators.

### 3. Graphene nano-ribbon transport model

The current along the channel can be calculated from Landauer's formula assuming a one-dimensional ballistic channel in between contacts that are further connected to external reservoirs, where dissipation takes place:

$$I = \frac{2q}{h} \sum_n \int_{-\infty}^{\infty} \text{sgn}(E) T_n(E) \big( f(\text{sgn}(E)(E - E_{FS})) - f(\text{sgn}(E)(E - E_{FD})) \big) dE \quad (4)$$

where $n$ is a natural number labeling subbands, $f(E)$ is the Fermi-Dirac distribution function, $T_n$ the transmission probability of the $n^{th}$-subband, and *sgn* refers to the sign function. This expression accounts for the spin degeneracy of the injected carriers. The current carried by each subband has been splitted into tunneling and thermionic components for carriers injected through and above the barrier respectively. Assuming phase-incoherent transport, transmission probabilities are computed through the S/D regions separately and then combined by using $T(E) = \frac{T_S T_D}{T_S + T_D + T_S T_D}$ in order to obtain the global transmission entering Landauer's formula [14]. Tunneling transmission probability through a single SB is computed using the Wentzel-Kramers-Brillouin (WKB) approximation $T(E) = \exp\left( -2 \int_{z_i}^{z_f} k(z) dz \right)$, where the wavevector k(z) is related with the energy by the GNR dispersion relation:

$$\pm \big( |E_{C,V}(z)| - |E| \big) + n \frac{E_g}{2} = \hbar v_F k(z), \quad (5)$$

where $v_F \sim 10^6$ m/s corresponds to the Fermi velocity of graphene, and the +/- sign applies to the calculation of tunneling and thermionic currents respectively. The integration limits appearing in the transmission formula are the classical turning points. For computing tunneling transmission close to the source contact note that, as long as $|\varphi_S|<E_g$, the turning points satisfy $z_i=0$ and $E_{C,V}(z_f)=E$ (for conduction and valence band respectively). In case of $|\varphi_S|>E_g$, the spatial band diagram curvature becomes high enough to trigger band-to-band tunneling (BTBT), and the turning points satisfy instead: $E_V(z_i)=E$ and $E_C(z_f)=E$ for electron BTBT; $E_C(z_i)=E$ and $E_V(z_f)=E$ for hole BTBT. Similar considerations must be done for tunneling through the drain contact barrier but replacing $\varphi_S$ by $\varphi_S-V_{DS}$. For energies $|E|$ above SB the thermionic transmission probability can be computed using WKB approach to yield [15]:

$$T(E) = \frac{16 k_C k_{GNR}^3}{\left(k'_{GNR}\right)^2 + 4\left(k_{GNR}^2 + k_C k_{GNR}\right)^2} \quad (6),$$

where $k_C$, $k_{GNR}$, are the wavevectors in the contact and the GNR region close to the contact respectively; the primed notation denotes a derivative respect to z. Assuming graphene metallic contacts $k_C = \frac{|E|-E_F}{\hbar v_F}$, with $E_F=E_{FS}=0$ at the source contact and $E_F=E_{FD}=-qV_{DS}$ at the drain contact. The $k_{GNR}$ wavevector at the S/GNR (D/GNR) interface can be easily obtained from Eqs. (1) and (5) at z=0 (z=L). Using the approximation $E_C(z) = SB - \varphi_S\left(1-e^{\frac{-\sqrt{2}z}{t_{ox}}}\right)$, the derivative of k(z) along z-direction yields

$$\left|k'_{GNR}\right| = \left|\frac{dE_{C,V}(z)/dz}{\hbar v_F}\right| \approx \frac{\sqrt{2}|\varphi_S|}{\hbar v_F t_{ins}} \quad (7),$$

for the S/GNR interface. The same expression holds for the D/GNR interface replacing $\varphi_S$ by $\varphi_S-V_{DS}$.

## 4. Model assessment

To assess the presented model I have simulated the same nominal device as used in Ref. [9]. It is formed by an armchair edge GNR channel with a ribbon index N=12, presenting a width $W = \sqrt{3}d_{cc}(N-1)/2 \approx$ 1.35 nm, where $d_{CC}$=0.142 nm refers to the carbon-carbon bond distance. Room temperature and bandgap of 0.83 eV were assumed for comparison purposes with the NEGF's method [9]. This value was estimated using tight-binding methods but a different bandgap $E_g \approx$ 0.6 eV results from a first-principles approach [12]. A gate insulator thickness $t_{ins}$=2 nm has been assumed. Note that the model, based on Laplace's equation, gives results not depending on the dielectric constant. The metal S/D is directly attached to the GNR channel, and SB height for both electrons and holes between the S/D and the channel is supposed to be half of the GNR bandgap $\varphi_{SB}=E_g/2$. The flatband voltage is zero. A power supply of $V_{DS}$=0.5 V has been assumed. The nominal device parameters have been varied to explore different scaling issues. The transfer characteristics exhibit two branches on the left and right from the minimum off-state current (Fig. 2). This minimum occurs at $V_{GS}=V_{DS}/2$ for a half-gap SB height, being the spatial band diagram symmetric for electrons and holes, and the respective currents are identical. This bias point is named the ambipolar conduction point. When $V_{GS}$ is greater (smaller) than $V_{DS}/2$, the SB width for electrons (holes) is reduced, producing a dominant electron (hole) tunneling current. The effect of power supply up-scaling is to further reduce SB width at the drain side making it more transparent and allowing more turn-on current to flow. The output characteristics of the SB GNR-FET are shown in the inset of Fig. 2, with an overestimation of the current in a factor about 2 respect to the NEGF's based model. The dominant current for the nominal device is electron tunneling and exhibits linear and saturation regimes. Increasing $V_{GS}$ produces a larger saturation current and voltage due to further transparency of SB and the expansion of the energy window for carrier injection from the source into the channel. Besides, downsizing W increases the gap and hence $\varphi_{SB}$ in the simulation (assumed to be $E_g/2$) further reducing the current due to a less populated higher energy levels [Fig. 3(a)], but the resulting on-off current ratio, a figure-of-merit for digital circuits, is largely improved. Reducing SB height respect to the half-gap case favors electron transport and results in a parallel shift of the ambipolar conduction point towards smaller gate voltages and asymmetries between the left and right branches of the transfer characteristic [Fig. 3(b)]. Also note that for low $\varphi_{SB}$ and $V_{GS}$, the thermionic electron current exceeds the tunneling electron current and should be taken into account for computing the off-state current. It is worth pointing out that for the thin insulator considered here the SB, which thickness is roughly the insulator gate thickness, is nearly transparent, producing a small effect on the qualitative feature of the transfer characteristics (only a parallel shift). Hence, it does not seem feasible to further reduce the off-state current by engineering the SB height. The scaling of gate insulator thickness improves gate electrostatic control producing larger transconductances and smaller subthreshold swings, as shown in Fig. 4. Also note that a thinner oxide produces a larger on-current and on-off current ratio. All results shown in Figs. 2-4 are in close agreement with that obtained with the NEGF's method despite I assumed a double gate geometry for the simulations presented in Figs. 2-3 instead of the single gate geometry from Ref. [9]. This observation points out the limited influence of the gate geometry for a quantum capacitance controlled device.

## 4. Conclusions

In conclusion, a simple model for the I-V characteristics of Schottky-barrier graphene field effect transistors which captures the main physical effects governing the operation of this device has been presented. Typical simulation of a I-V characteristic with 100 points takes no more than few seconds on a personal computer. The results obtained applying this model to prototype devices are in close agreement with a more rigorous treatment based on the NEGF's approach, thus validating the approximations made.

The presented model is intended to assist at the design stage as well as for quantitative understanding of experiments involving GNR-FETs.

Financial support of this work was provided by Ministerio de Educación y Ciencia under project TEC2006-13731-C02-01/MIC.

**References**

[1] Geim A K and Novoselov K S 2007 *Nature Materials* **6** 183

[2] Novoselov K S and Firsov A A 2004 *Science* **306** 666

[3] Berger C, Song Z, Li X, Wu W, Brown N, Naud C, Mayou D, Li T, Hass J, Marchenkov A N, Conrad E H, First P N and de Heer W A 2006 *Science* **312** 1191

[4] Avouris P, Chen Z and Perebeinos V 2007 *Nature Nanotech.* **2** 605

[5] Nakada K, Fujita M, Dresselhaus G and Dresselhaus M S 1996 *Phys. Rev.* B **54** 17954

[6] Obradovic B, Kotlyar R, Heinz F, Matagne P, Rakshit T, Giles M D, Stettler M A and Nikonov D E 2006 *Appl. Phys. Lett.* **88** 142102

[7] Jiménez D, Cartoixà X, Miranda E, Suñé J, Chaves F A and Roche S 2007 *Nanotechnology* **18** 025201

[8] Fedorov G, Tselev A, Jiménez D, Latil S, Kalugin N G, Barbara P, Smirnov D and Roche S 2007 *Nano Lett.* **7** 960

[9] Ouyang Y, Yoon Y and Guo J 2007 *IEEE Trans Electron Devices* **54** 2223

[10] Guo J, Yoon Y and Ouyang Y 2007 *Nano Lett.* **7** 1935

[11] Morse P M and Feshbach H, *Methods of Theoretical Physics (*McGraw-Hill, New York, 1953), Chap. 10, p. 1247.

[12] Son Y W, Cohen M L and Louis S G 2006 *Phys. Rev. Lett.* **97** 216803

[13] Datta S, *Quantum transport: atom to transistor* (Cambrigde University Press, Cambridge, 2005), Chap. 7, pp. 172-173.

[14] Datta S, *Electronic transport in mesoscopic systems* (Cambrigde University Press, Cambridge, 1995), Chap. 2, p. 63.

[15] John D L, *Simulation studies of carbon nanotube field-effect transistors* (PhD thesis, University of British Columbia, 2006), Chap 4, p. 23.

**Figure captions**

**Figure 1.** Geometry and band diagram of the GNR-FET: (a) cross-section, (b) top view of the armchair shaped edge GNR forming the channel, and (c) sketch of the spatial band diagram along transport direction.

**Figure 2.** Transfer and output characteristics (inset) for the nominal GNR-FET. Decomposition of the total current in electron and hole tunneling contributions are shown.

**Figure 3.** Influence of the GNR width (a) and SB height (b) in the transfer characteristics.

**Figure 4.** Impact of the gate insulator thickness scaling on the transfer characteristics. The inset shows the effect of scaling on the transconductance for $V_{GS}$=0.75 V and subthreshold swing.

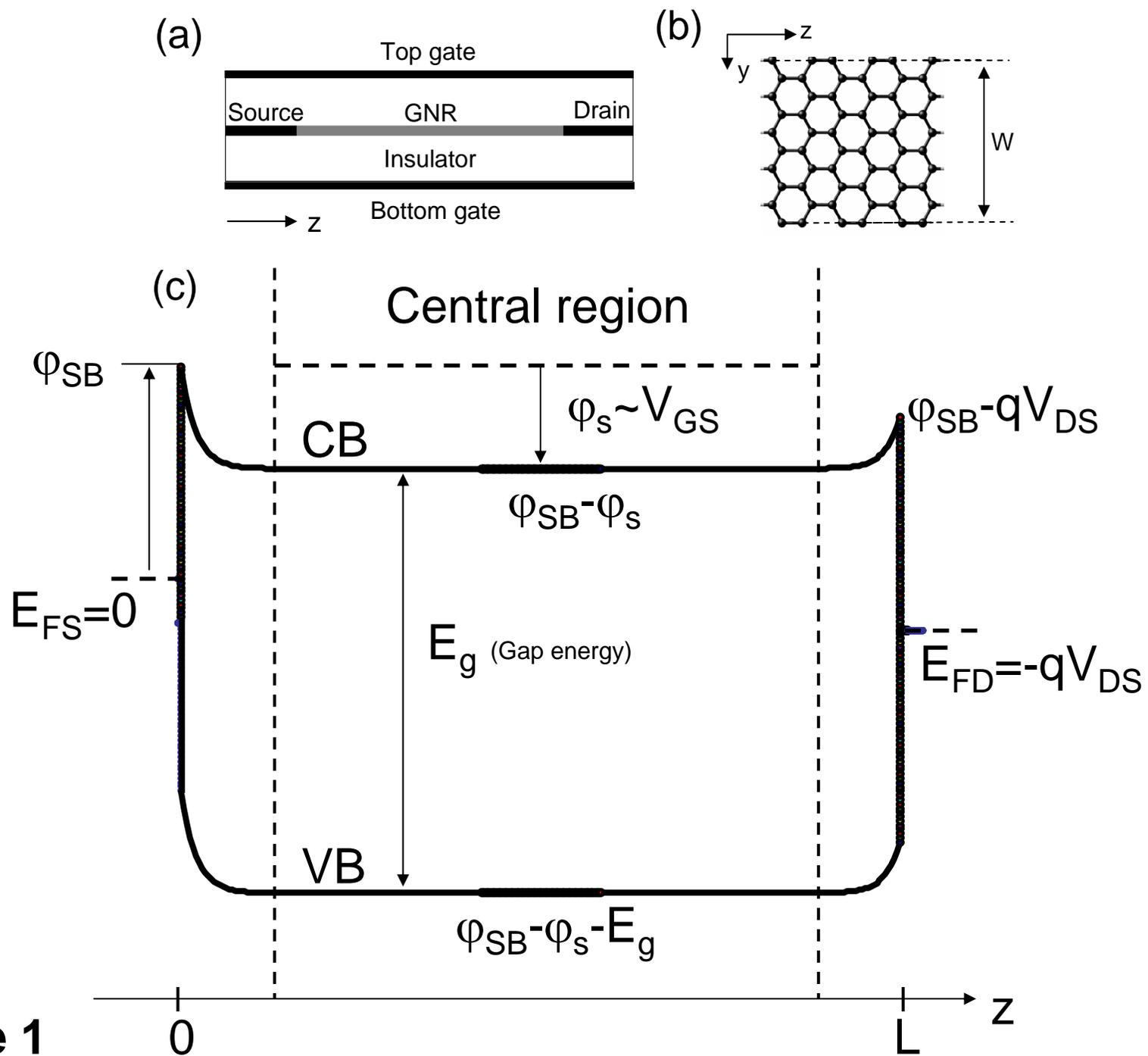

Figure 1

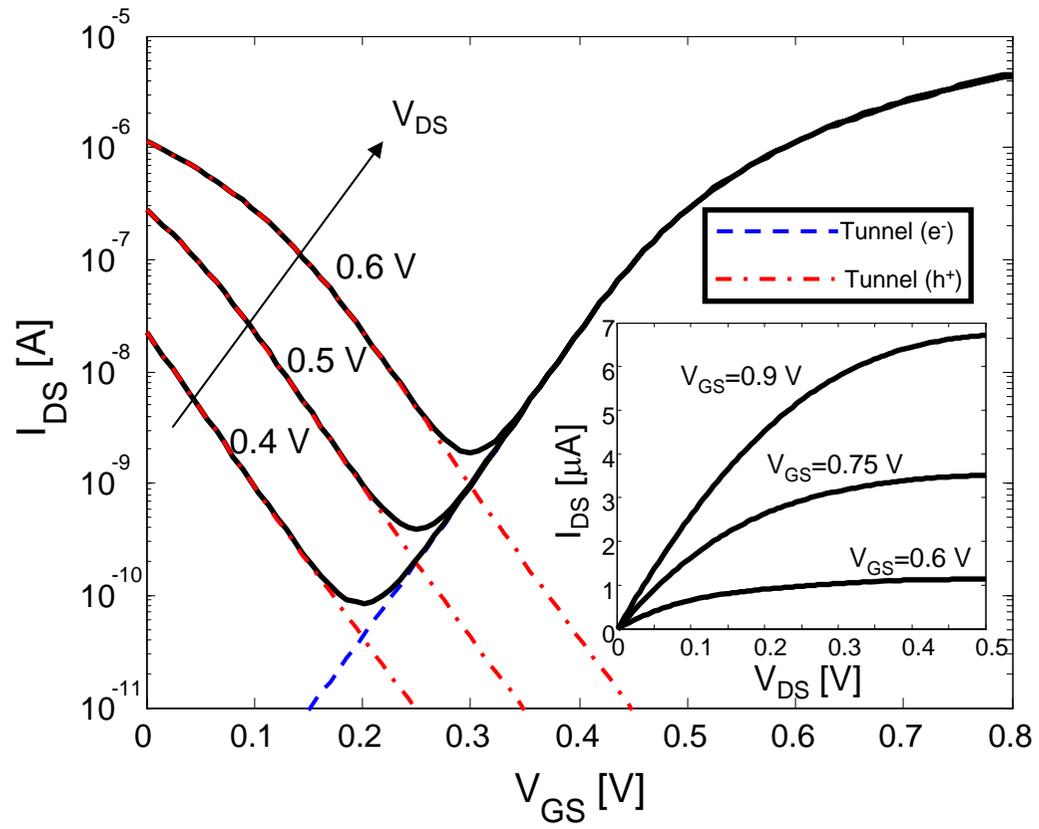

**Figure 2**

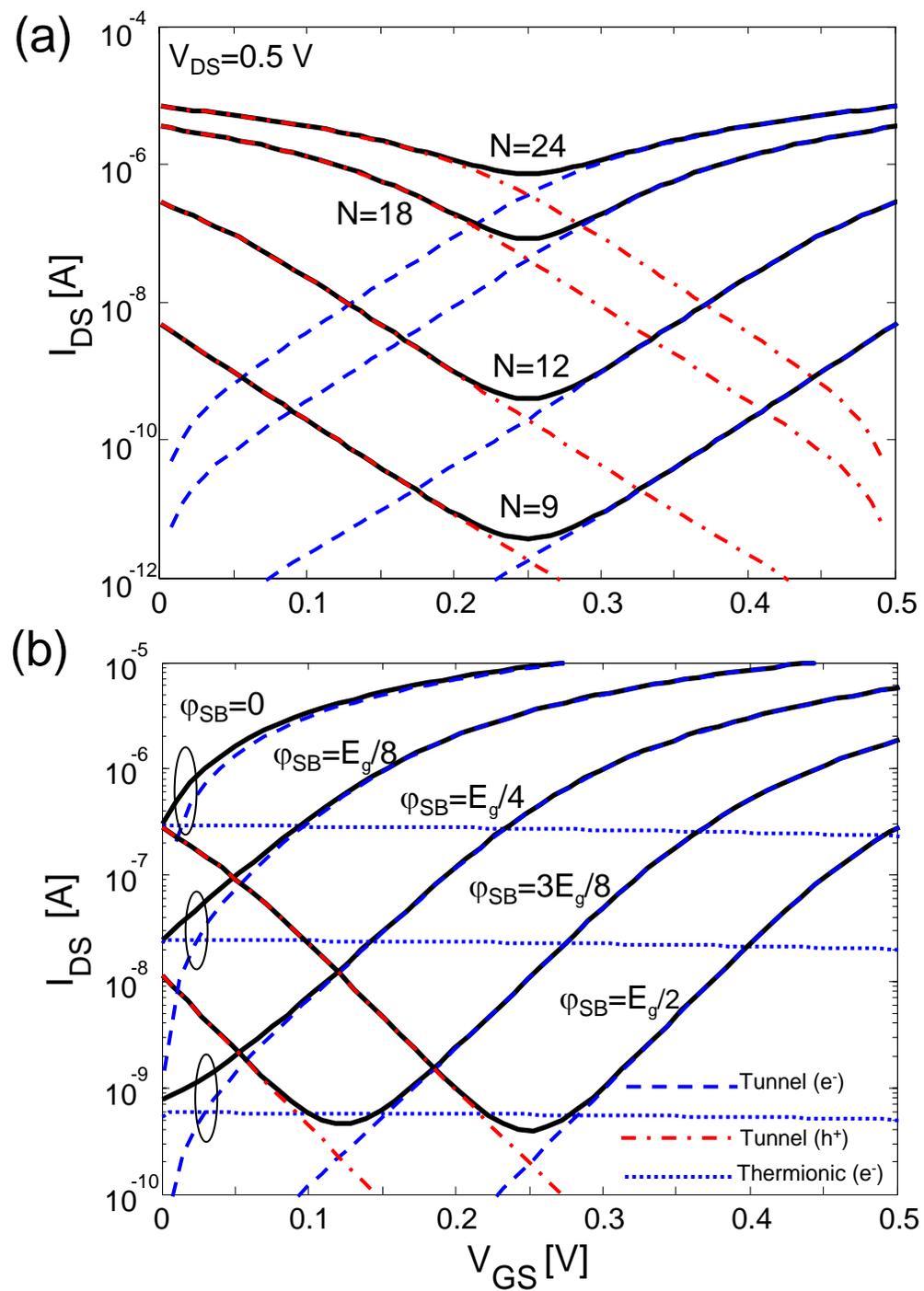

Figure 3

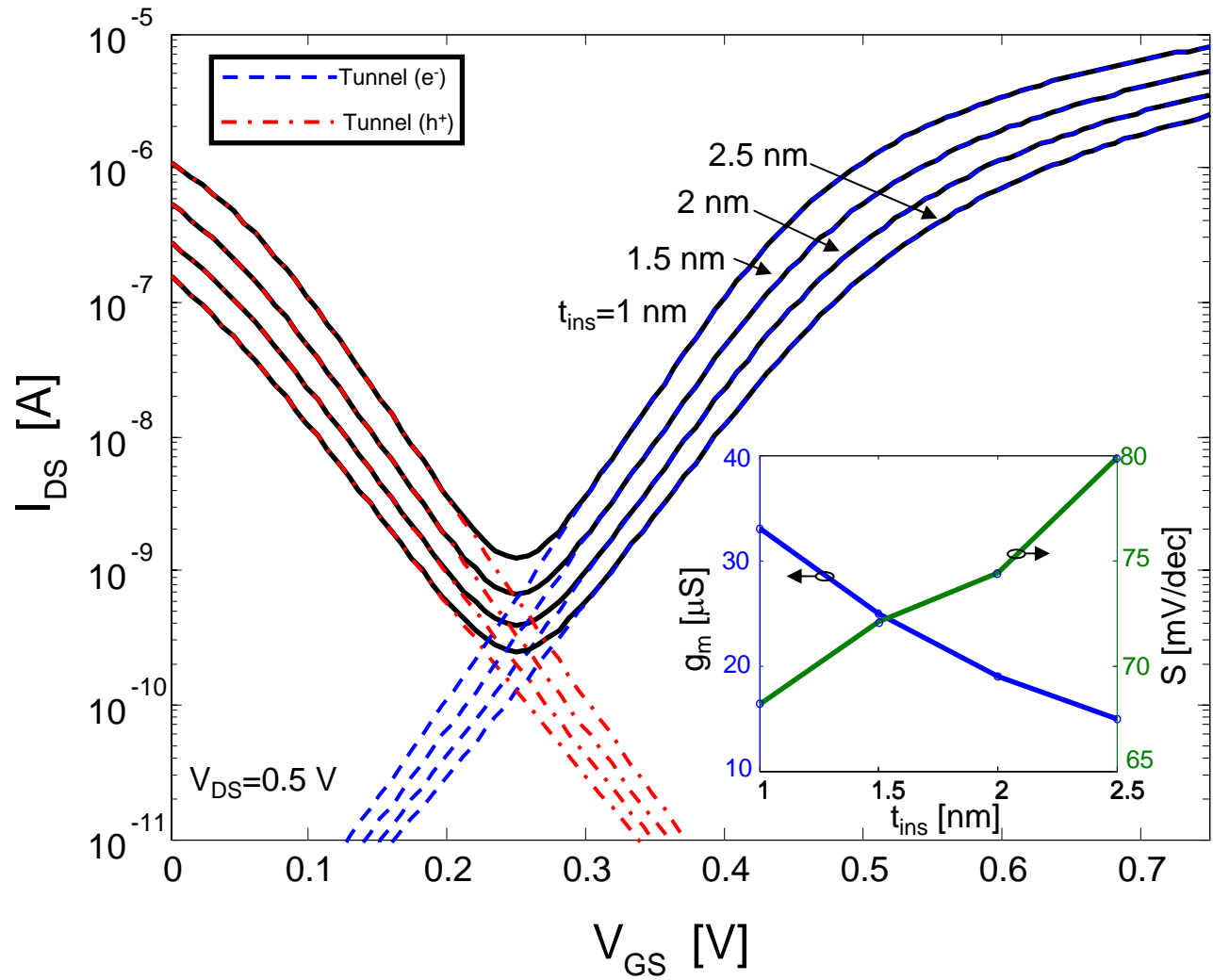

Figure 4